\documentclass[aps,11pts, twocolumn,notitlepage,longbibliography]{revtex4-1}

\usepackage{graphicx}
\usepackage[usenames]{color}
\usepackage{amsmath,amssymb}
\usepackage{gensymb}
\usepackage{natbib}
\usepackage{tabularx}
\usepackage{longtable}
\usepackage{xcolor}
\usepackage{upgreek}
\usepackage{hyperref}
\usepackage{subfigure}
\usepackage{changes}
\usepackage{upgreek}
\usepackage{changes}
\newcommand{\bs}{\boldsymbol}

\usepackage{multirow}

\usepackage{bbm}

\begin{document}
\title{Controlling transport dynamics of confined asymmetric fibers}

\author{M. Bechert$^{1}$}\thanks{These authors contributed equally to this work.} \author{J. Cappello$^{2}$,${^*}$, M. Da\"{\i}eff$^{2}$, F. Gallaire$^{1}$, A. Lindner$^{2}$}\thanks{anke.lindner@espci.fr} \author{C. Duprat$^{3}$}



\affiliation{                    
  $^{1}$ Laboratory of Fluid Mechanics and Instabilities, \'Ecole Polytechnique F\'ed\'erale de Lausanne, Lausanne 1015, Switzerland\\
  $^{2}$ Laboratoire de Physique et M\'ecanique des Milieux H\'et\'erog\`enes, UMR 7636, CNRS, ESPCI Paris, PSL Research University, Universit\'e Paris Diderot, Sorbonne Universit\'e, Paris, 75005, France\\
  $^{3}$ Laboratoire d'Hydrodynamique (LadHyX), \'Ecole polytechnique, Department of Mechanics, Palaiseau, France\\
  }

\begin{abstract}
{Transport properties of particles in confining geometries show very specific characteristics as lateral drift, oscillatory movement between lateral walls or the deformation of flexible fibers. These dynamics result from viscous friction with transversal and lateral channel walls inducing strong flow perturbations around the particles that act like moving obstacles.  In this paper, we modify the fiber shape by adding an additional, small fiber arm, which leads to T and L shaped fibers with only one or, respectively, zero symmetry axes and investigate the transport properties. For this purpose, we combine precise microfluidic experiments and numerical simulations based on modified Brinkman equations. Even for small shape perturbations, the transport dynamics change fundamentally and formerly stable configurations become unstable, leading to non-monotonous fiber rotation and lateral drift. Our results show that the fundamental transport dynamics change with respect to the level of fiber symmetry, which thus enables a precise control of particle trajectories and which can further be used for targeted delivery, particle sorting or capture inside microchannels.}
\end{abstract}


\maketitle

\section{Introduction}
Separating and filtering of particles or micro-organisms as a function of their properties such as flexibility, shape or activity is an important requirement for biomedical or food science applications. Micronscale filters, including micro-pillar arrays 
\cite{Gossett2010, DiCarlo2009} are commonly used to sort particles with respect to their size or deformability, but face problems of filter clogging or damaging of deformable particles. Recently microfluidic devices, relying on specific transport properties, have been developed to overcome these difficulties \cite{Gossett2010}. Inertial \cite{Masaeli2012, DiCarlo2009} or viscoelastic effects \cite{DAvino2017} can be used to focus particles at specific positions inside channels and flexible particles are known to migrate away from bounding walls, leading for example to a cell free layer as observed in blood flow \cite{Secomb2017} or migration of flexible fibers in shear gradients \cite{duRoure2019}. Some microorganisms use passive reorientation mechanisms to move in gradients, either in the gravitational field (gravitaxis)\cite{TenHagen2014} or in a gradient of viscosity (viscotaxis)\cite{liebchen2018}. In simple unbounded flows however, when inertia can be neglected, rigid particles only migrate across streamlines when symmetry is broken for example by particle chirality\cite{Marcos2009}. 

\begin{figure}
  \begin{center}
\includegraphics[width=0.99 \columnwidth]{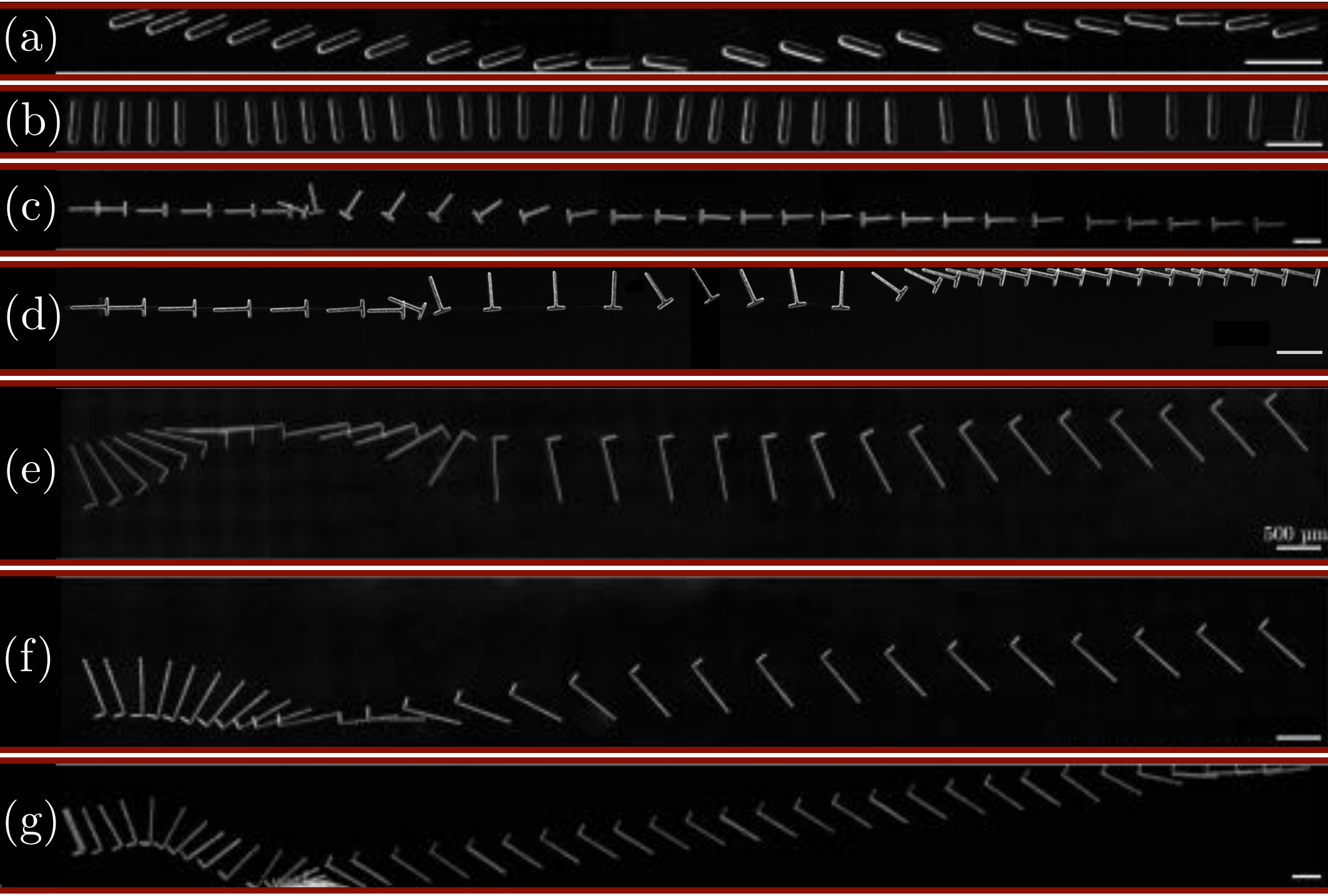}
  \end{center}
\caption{{Experimental} chronophotographies of transported fibers of different geometries. The fluid flows from left to right.{ The lateral walls are highlight in red.} As reported previously, straight fibers oscillate in the channel either through glancing (a) or reversing (b) \cite{nagel2018}. T-fibers away from the lateral walls reorient toward a stable orientation parallel to the flow direction (c) {or are captured by the wall (d)}. L-fibers first rotate toward a stable orientation and then drift {(e-g)} until they are captured by the lateral wall ({g}). From top to bottom confinements are (a,b) 0.78, (c{,d}) 0.82, ({e-f}) 0.76, and ({g}) 0.72. Scale bars are 500 $\upmu$m.{Total length of the channels is typically 4 cm. The time interval between successive snapshots is constant for each panel and varies between $1\,$s and $7\,$s (such that the fiber is translated about its length between successive images).}}
\label{fig:reorientation}
\end{figure}

This situation changes when particles are confined by bounding walls and act as moving obstacles to the flow. The induced strong flow perturbation leads to particle migration, as for example lateral drift observed for rigid fibers even at small Reynolds numbers \cite{Berthet2013}. Recently, theoretical models and experiments have been developed to describe such flows \cite{Berthet2013,uspal2013,Bet2018b,shen2014,nagel2018}. In particular, it has been shown that axisymmetric objects such as fibers or symmetric dumbbells (made of rigid spheres or drops) translate without rotation but drift at a constant angle, and may oscillate between the lateral walls of the channel \cite{nagel2018} (Fig.~\ref{fig:reorientation}(a,b)). Axisymmetric particles with fore-aft asymmetry, such as asymmetric dumbbells, will migrate toward the center of the channel, where they align with the flow \cite{uspal2013, shen2014}, while laterally unconfined, asymmetric trimers rotate to reach an equilibrium angle \cite{Bet2018}. Flexible fibers may be deformed by the viscous forces and will then rotate and align with the flow \cite{Cappello2019, duRoure2019}. In all these cases, the particle trajectory is strongly affected by the transversal confinement, which tunes the magnitude and distribution of the viscous force.

This situation bears some similarities with sedimenting particles where, in particular, there exists a coupling between translation and rotation that depends on the particle geometry, and particles that possess certain symmetries will exhibit specific motions. For example, axially symmetric objects, such as uniform rods, keep their orientation and merely translate without rotating \cite{Cox1970}; an axisymmetric object that presents a fore/aft asymmetry, such as a dumbbell composed of spheres of different sizes, will rotate and align with the flow \cite{Candelier2016}. An asymmetric particle, as for example an L shaped particle, will rotate until it reaches a stable orientation, at which it will translate without rotating\cite{TenHagen2014}. If in addition the chiral symmetry is broken, the coupling of translational and rotational movement leads to helical trajectories \cite{Tozzi2011, Palusa2018}.

We investigate these effects in detail by studying a model system consisting of a microfiber with increasing degrees of asymmetry transported in a microchannel. By adding a second arm to an initially straight fiber, we create and analyze, both theoretically and experimentally, T-shaped fibers with fore/aft asymmetry and fully asymmetric L-shaped fibers. Examples of these fibers and corresponding chronophotographies of their transport dynamics are given by Fig.~\ref{fig:reorientation}(c-{g}).\par

Straight fibers, which we will refer to as ``I fibers" throughout this paper, oscillate between the lateral channel walls while transported downstream  (Fig.~\ref{fig:reorientation}(a,b)). {The fiber rotation accompanying the oscillatory movement results from interaction with the lateral walls, i.e., away from these walls, the fiber does not rotate and merely drifts.} In contrast the interplay of the long and short arms of the T and L fibers induces rotation even away from the lateral walls. T fibers rotate in most cases until aligned with the flow, with the small branch at the tail, and remain parallel to the flow while slowly migrating toward the center of the channel (Fig.~\ref{fig:reorientation}(c)). {In some cases, due to the lateral confinement, the fiber is captured by the wall while it reorients toward the aligned position (Fig.~\ref{fig:reorientation}(d))}. L fibers rotate, either clockwise or anti-clockwise, until they reach an equilibrium angle which they keep while drifting toward the lateral wall (Fig.~\ref{fig:reorientation}({e},{f})). At the wall, two different scenarii can be observed. Fibers arriving with the sharp edge ``rebound" off the wall, i.e. rotate until reaching their equilibrium angle and drift toward the opposite wall. Fibers arriving with the open side impact the wall and remain stuck at the wall (Fig.~\ref{fig:reorientation}({e-g})).\par

In the following, we will primarily focus on L shaped fibers, being the most asymmetric objects, and complete the picture at the end of this letter by full trajectory diagrams in the configuration space of I, T and L fibers, revealing the fundamental difference between these three types of fibers and symmetries.

\section{Problem formulation and methods}

\begin{figure}
 \begin{center}
\includegraphics[width=1\columnwidth]{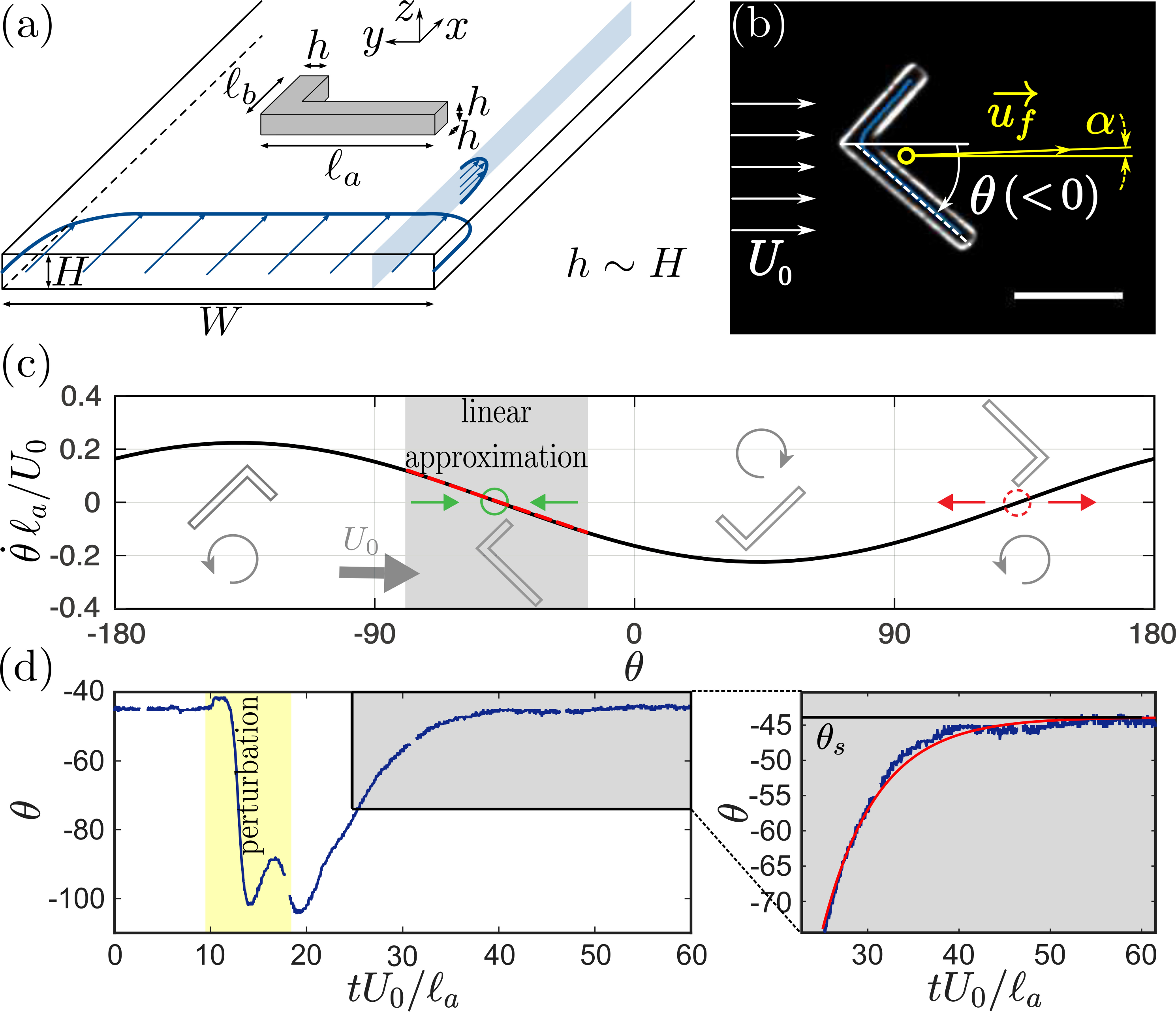} 
 \end{center}
 \caption{(a) Geometry of an L-shaped fiber in the channel. The pressure-driven flow is sketched in blue. (b) Image treatment and definition of angles. Using MatLab procedures we derive from the picture of the particle its skeletonized shape (blue) and its center of mass (yellow circle). The scale bar is 200 $\upmu$m. (c) Evolution of the dimensionless rotation velocity $\dot{\theta}\,\ell_a/U_0$ with the angle $\theta$ for an L fiber with $\ell_a/h = 10$ and $\ell_b/h = 5$ for $\beta=0.76$, far away from the lateral walls. The green circle indicates a stable fixed point and the red circle an unstable fixed point. (d) Experimentally observed evolution of the orientation angle $\theta$ as a function of the dimensionless time (dark blue). A fiber initially in its equilibrium orientation hits an obstacle (yellow stripe) and deviates from the equilibrium orientation. It then rotates back to the latter. The reorientation dynamics close to the equilibrium orientation (grey area) are well adjusted by an exponential fit (red curve in the zoom). The fitting method is detailed in appendix \ref{app:exp_measure}.}
\label{fig:channel_fiber_geom}
\end{figure}

Figures~\ref{fig:channel_fiber_geom}~(a,b) depict the configuration of an L fiber in the channel. The fiber is of square cross-section with width and height $h$, and $\ell_a$ and $\ell_b$ denote the lengths of the long and short arms, respectively. We will refer to $\ell_a/h$ and $\ell_b/h$ as the non-dimensional lengths or aspect ratios of the long and short fiber arms, respectively, and quantify the asymmetry by $\ell_a/\ell_b$, which is bounded by unity ($\ell_b=\ell_a$, symmetric fiber with two arms of equal length) and $\ell_a/h$ ($\ell_b=h$, I fiber). Starting from $\ell_a=\ell_b$, the asymmetry is first enhanced when decreasing the length of the short arm with respect to the long arm before the symmetry is restored when the short arm vanishes and a straight (I) fiber is attained.

The fiber is placed in a rectangular channel of height $H$ and width $W$, and subjected to a pressure driven flow fully characterized by its mean velocity $U_0$. {Due to the Hele-Shaw-like geometry of the channel (the aspect ratio $W/H$ of the channel varies from 15 to 70), the flow in the $xy$ plane is a plug flow, except close to the side walls and in the vicinity of the fiber, where the transverse vorticity is confined within boundary layers of characteristic thickness $H$. In the $z$ direction the flow is Poiseuille-like. }The transversal and lateral confinements of the fiber are quantified by $\beta=h/H$ and $\xi=\ell_a/W$, respectively.  As the channel is assumed to be of infinite length, the state of the fiber is completely described by the lateral position of its center of mass $y$ and the orientation angle $\theta$, together with the corresponding velocities $\dot y$ and $\dot\theta$. Moreover we define the drift angle $\alpha$ by $\tan\alpha = \dot y/\dot x$. Note that there exists  a mirror symmetry and that the mirror image of the fiber shown in Fig.~\ref{fig:channel_fiber_geom}(b) corresponds to opposite fiber chirality.  \par

The experimental methods, theoretical model, and numerical techniques are extensively described and verified in \cite{nagel2018} and \cite{Cappello2019} and are outlined only briefly here. Fibers of controlled shape, position and confinement are fabricated using the stop-flow microscope-based projection photo-lithography method developed by Dendukuri \textit{et al.} \cite{Dendukuri2007}: a pulse of UV-light passing through a mask illuminates the microchannel filled with a photosensitive solution (PEG-DA Mw=575 and Darocur 1173, viscosity  $\mu=68\, \rm mPa\,\rm s$) to create a rigid particle which is then transported in the non-crosslinked solution, with a pressure-driven flow imposed by a syringe pump (Nemesys, Cetoni). The presence of inhibition, i.e. uncured, layers of constant thickness $b = 6 \pm 1.6\,\upmu$m in the vicinity of the channel walls \cite{Dendukuri2008} determines the fiber height $h = H-2b$, and thus the transversal confinement $\beta$. For each channel height, we adjust the mask geometry to ensure a square cross-section of side $h$ and a constant aspect ratio $\ell_b/h$ = 5. With this method, the asymmetry $\ell_a/\ell_b$ is tuned by changing the length of the long arm $\ell_a/h$. {Due to the low value of the Reynolds number $Re\sim10^{-3}$, inertia will be neglected in this work. This also implies that the dynamics does not explicitly depend on the flow velocity.}\par

A two-dimensional Brinkman model supplemented by a model flow profile in the gap between the fiber and the transversal channel walls is utilized to determine the flow around the fiber and the resulting force and torque distribution on the fiber. Note that the velocity vector is uniquely determined by the particular state $(y,\theta)$, as inertial effects are negligible and the problem is thus reversible in time. The flow equations are solved with \textsc{Comsol Multiphysics} and verified by separate calculations with the \textsc{Ulambator} code \cite{Nagel2015}. {Detailed information on the model can be found in previous works \cite{nagel2018, Cappello2019} and in the appendix \ref{app:Brinkman}.}

\section{Results}

We first focus on the equilibrium position reached by the fiber far from the lateral walls, i.e. $\xi\rightarrow0$. While for an I fiber all orientations are stable,  i.e. the fiber keeps a constant orientation and translates without rotating, the L fiber reorients toward a stable orientation, and then drifts at a constant angle (Figs.~\ref{fig:reorientation}({e-f})). We obtain numerically the evolution of the rotation speed $\dot\theta$ as a function of $\theta$ for an asymmetric L fiber with $\ell_a/h=10$, $\ell_b/h=5$, and $\beta=0.76$ (Fig.~\ref{fig:channel_fiber_geom}(c)). There are two fix points separated by $180^\circ$: a stable one {($\theta = \theta_s=-47^\circ$)} and an unstable one {($\theta = \theta_u=133^\circ$)}. For all initial orientations, the fiber rotates monotonously until the stable orientation $\theta_s$ is reached. Note that close to the unstable orientation $\theta_u$, a small variation can change the direction of rotation as depicted in Figs.~\ref{fig:reorientation} {(e,f)}; although both fibers start at very similar configurations ($\theta\simeq 130^\circ$), the first fiber is rotating counter-clockwise while the second is rotating in the opposite direction until they both reach the stable {orientation}. A fiber pushed out of the equilibrium orientation by an obstacle in the channel Fig.~\ref{fig:channel_fiber_geom}(d) rotates immediately back to the equilibrium position, showing the attractive strength of the stable fixed point. Close to the stable orientation, the rotation velocity $\dot \theta$ depends approximately linearly on $\theta$ (dashed line in Fig.~\ref{fig:channel_fiber_geom}(c)), i.e. the angle increases/decreases exponentially in time to reach its stable value $\theta_s$. We use this result to determine $\theta_s$ experimentally with an exponential fit (Fig.~\ref{fig:channel_fiber_geom}(d)), as detailed in the appendix \ref{app:exp_measure}.\par

\begin{figure}
  \begin{center}
    \includegraphics[width=0.8\columnwidth]{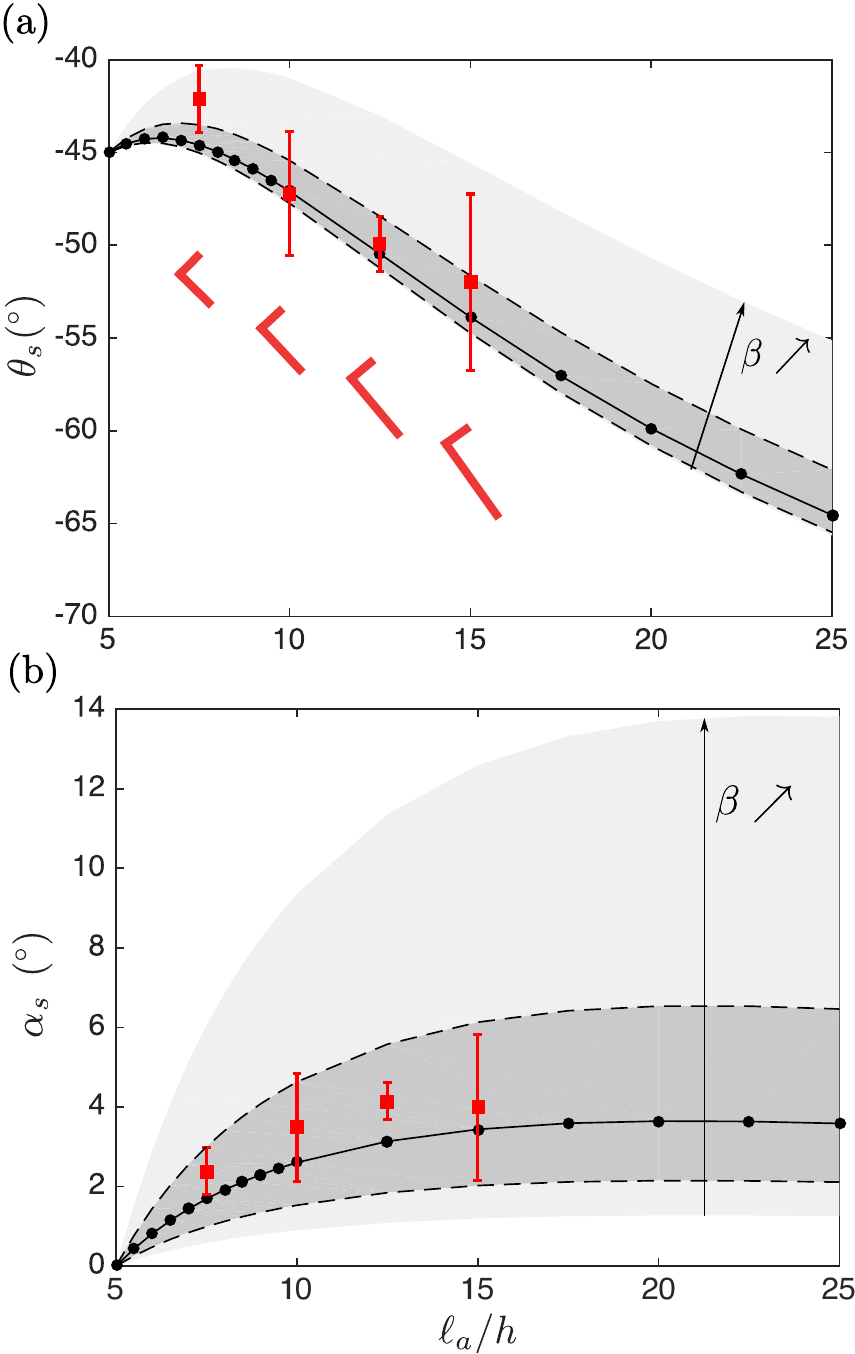}
  \end{center}
  \caption{Stable orientation angle $\theta_s$ (a) and corresponding stable drift angle $\alpha_s$ (b) as a function of the long arm aspect ratio $\ell_a/h$, for $\ell_b/h = 5$. Black squares show experimental measurements, red circles correspond to numerical calculation with a transversal confinement of $\beta =  0.76$. The two grey shades correspond to {the numerically determined variations for} $\beta=0.76\pm0.08$ and $\beta=0.76\pm0.16$ respectively. {The error bars correspond to the standard deviation of the measurements.}}
\label{fig:angle_eq_comp_exp_simu}
\end{figure}

The stable orientation angle $\theta_s$ and the corresponding drift angle $\alpha_s$ are uniquely defined by fiber geometry and transversal confinement. Figure~\ref{fig:angle_eq_comp_exp_simu} shows both quantities as determined in the experiment for different lengths of the long arm $\ell_a/h$, i.e. for varying asymmetry, and compared to the numerical results. While the drift angle varies only slightly in the examined regime, the stable orientation angle $\theta_s$ evolves non-monotonously with increasing length of the long arm $\ell_a/h$: it first increases, reaches a maximum at $\ell_a/h \sim 7.5$, and then decreases again. A variation of the transversal confinement impacts both $\theta_s$ and $\alpha_s$, as visualized by the gray shaded regions in Fig.~\ref{fig:angle_eq_comp_exp_simu}. While the effect of $\beta$ is almost negligible for $\beta<0.6$, the orientation and drift angles increase rapidly with increasing confinement. In particular, we found that increasing the transversal confinement increases the rotation speed $\dot{\theta}$, i.e., the higher (lower) the transversal confinement, the faster (slower) the fiber approaches the stable orientation (see appendix \ref{app:dynamics}).

\begin{figure}
  \begin{center}
 \includegraphics[width=\columnwidth]{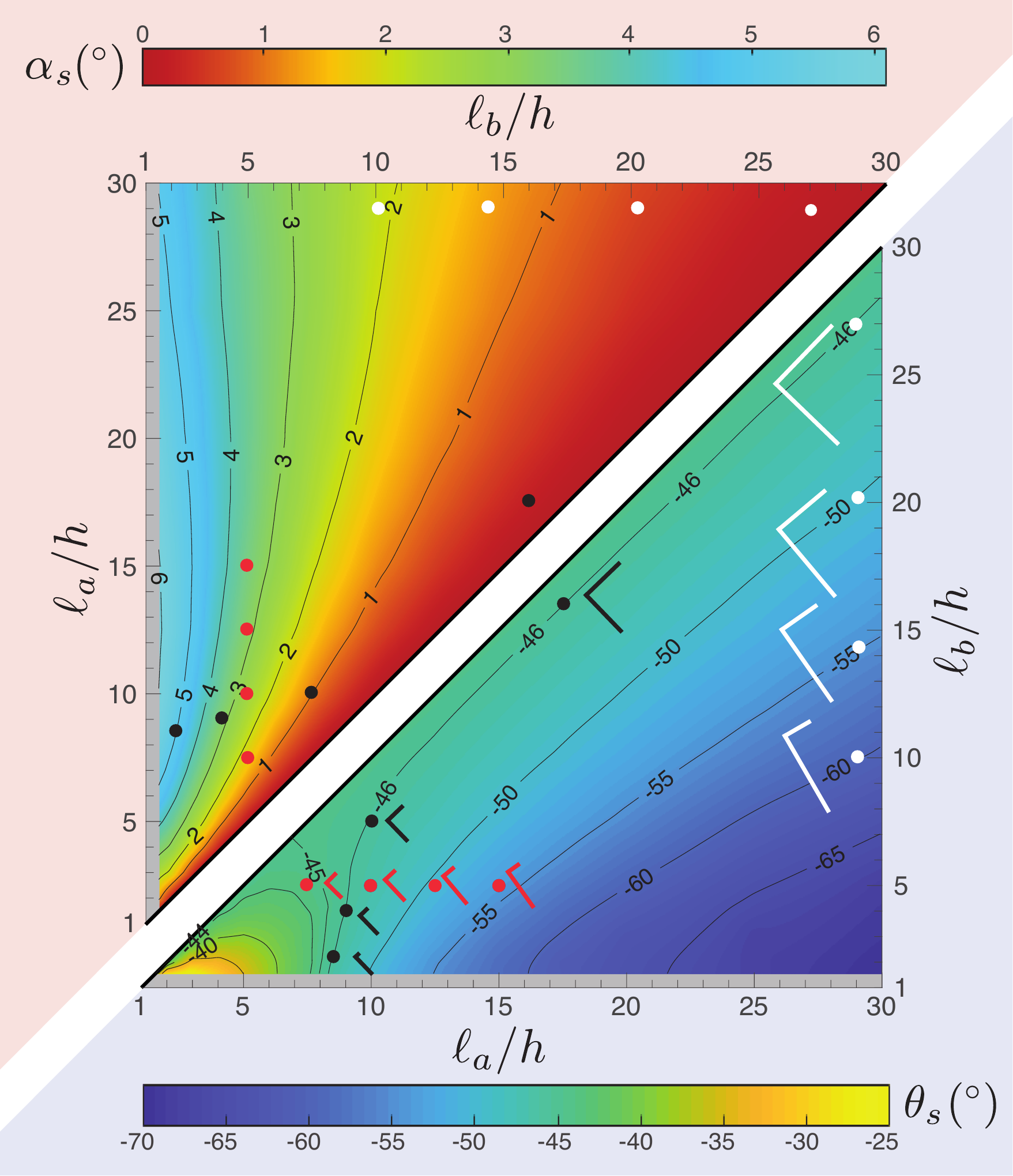}
 \end{center}
  \caption{{Numerically determined} isolines of constant stable orientation angle $\theta_s$ (bottom right corner) and constant stable drift angle $\alpha_s$ (top left corner) as a function of the fiber arm aspect ratios $\ell_a/h$ and $\ell_b/h$. Both maps are symmetric around the black line defined by $\ell_a=\ell_b$, where $\theta_s=45^\circ$ and $\alpha_s=0$. Numerically inaccessible regions are shaded in gray.}
\label{fig:equili}

\end{figure}

In order to obtain a general overview of the dependence of $\theta_s$ and $\alpha_s$ on the fiber geometry, we systematically compute these quantities in the parameter space $\{\ell_a/h$, $\ell_b/h\}$ and visualize the result in a map as shown in Fig.~\ref{fig:equili}. {Due to symmetry and for the sake of compactness, we restrict the space to $\ell_a\leq\ell_b$ and combine the two maps for $\alpha_s$~(top) and $\theta_s$~(bottom) in one. The black line corresponds to $\ell_a=\ell_b$, where the} fiber is symmetric and is transported with $\theta_s=-45^{\circ}$ along the flow direction, i.e. $\alpha_s=0^{\circ}$. For $\ell_a>\ell_b$ the L fiber is fully asymmetric as long as $\ell_b>h$, and the map of stable orientations $\theta_s$ can be divided into two distinct regions. For large values of the aspect ratios $\ell_a/h$ and $\ell_b/h$, the isolines of constant $\theta_s$ approach straight lines i.e. $\theta_s$ depends on the asymmetry ratio $\ell_a/\ell_b$ only and is independent of the fiber height $h$ and thus the fiber aspect ratios. In this region, increasing asymmetry ($\ell_a/\ell_b$ decreasing) leads to a decrease of $\theta_s$, as evidenced by the white dots and the corresponding fiber shapes.

At small values of the aspect ratios ($\leq 10$), the isolines exhibit a different tendency: changing $\ell_a$ or $\ell_b$ while keeping $\ell_a/\ell_b$ constant leads to different values of $\theta_s$, i.e $\theta_s$ depends not only on the fiber asymmetry, but also on the fiber aspect ratios. Keeping for example $\ell_b/h$ constant and changing $\ell_a/h$ (red dots in Fig.~\ref{fig:equili}), as we did in the experiments shown in Fig.~\ref{fig:angle_eq_comp_exp_simu}, leads to a non-monotonous evolution of the orientation angle. The crossover between these two regimes can be evidenced by following the isoline for $\theta_s=-46^{\circ}$ (black dots and corresponding shapes). Along this line, the fiber shape first changes significantly at small fiber arm aspect ratios, but finally reaches an homothetically similar shape at large values of the aspect ratio.\par

For asymmetric fibers, which form an angle $\theta_s$ with the flow direction, there is a lateral drift whose magnitude depends on a combination of both equilibrium angle and asymmetry. For a fixed $\ell_b/h$, changing $\ell_a/h$ (red dots) leads to a small change in drift angle although the orientation angle varies significantly. On the contrary, for a given orientation (black dots), changing the shape leads to a significant change in drift angle. In the limit of large fiber arm aspect ratios, changing the asymmetry $\ell_a/\ell_b$ (white dots)  leads to significant variations in orientation and drift angle. \par

These findings demonstrate that it is not possible in general to rationalize the fiber transport using only the fiber asymmetry $\ell_a/\ell_b$, but that the two-dimensional geometry of the fiber has to be taken into account instead of a one-dimensional skeleton.

\begin{figure*}
  \begin{center}
\includegraphics[width=\textwidth]{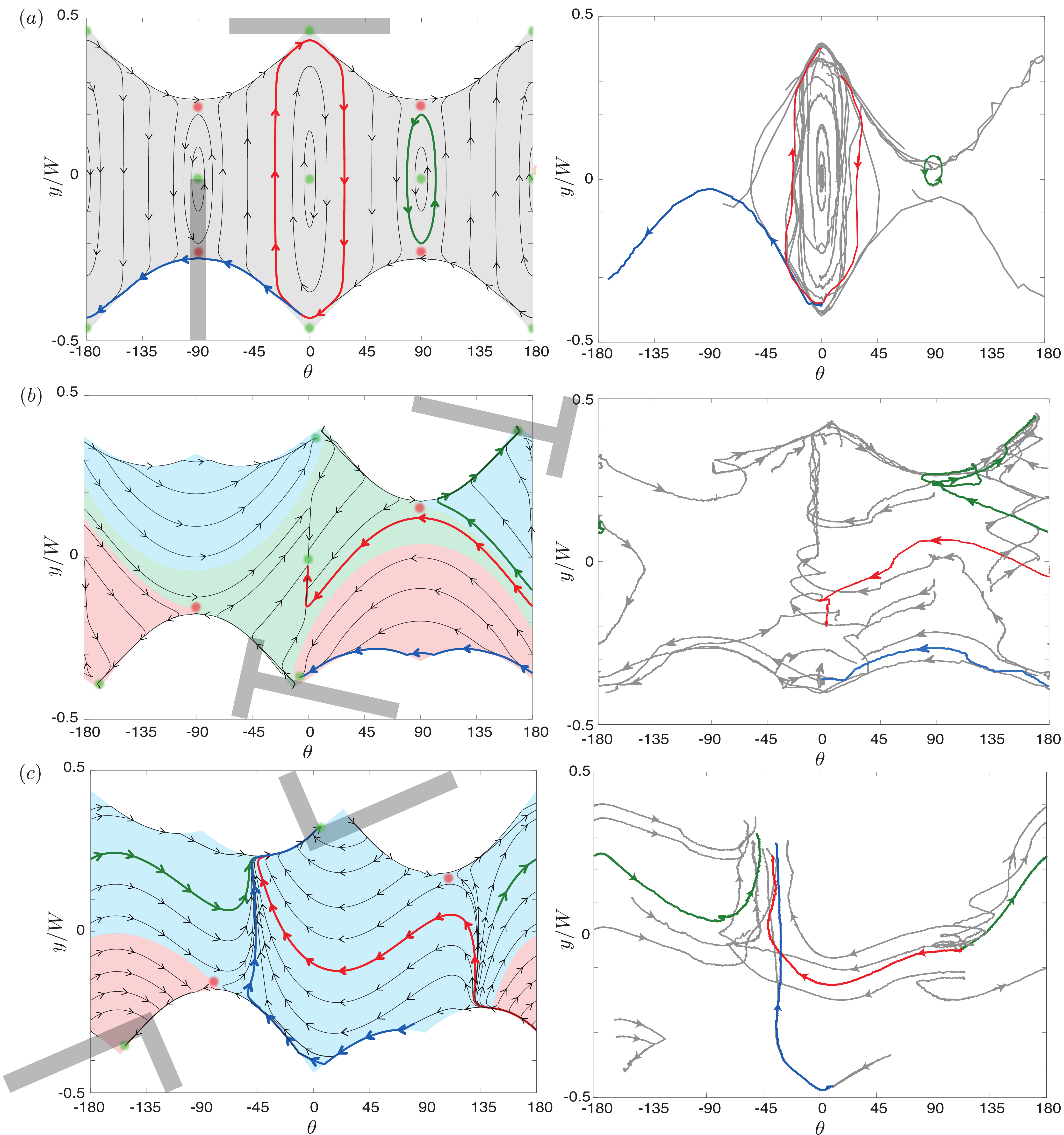}
 \end{center}
  \caption{Trajectory map in the configuration space spanned by orientation angle $\theta$ and lateral position $y$ for (a) an I fiber, (b) a T fiber, and (c) an L fiber {for numerics (left) and experiments (right)}. The physical accessible space is shaded in color. For T fibers, the space of trajectories ending at the channel center is shaded in light green. The space of trajectories ending at the wall at $y=0.5$ (resp. $y=-0.5$) is shaded in light blue (resp. light red). Stable and unstable fix points are highlighted in green and red, respectively. {Experiments: transversal confinement $0.7\leq \beta\leq0.8$, lateral confinement $0.12\leq\xi\leq0.8$, and arm lengths $10\leq\ell_a/h\leq20$, $\ell_b/h=5$. Numerics: Long arm aspect ratio $\ell_a/h= 10$, lateral confinement $\xi=0.5$ and transversal confinement $\beta=0.76$. The short arms of the T and L fibers have half the length of the long arm, i.e., $\ell_b/h=5$.} Some numerical and experimental trajectories are highlighted for comparison.}
\label{fig:fiber_trajectories}
\end{figure*}

The overall fiber dynamics is given by a combination of rotation toward $\theta_s$, drift, and interaction with the lateral walls. {For a more global view, we plot the trajectories in the configuration space spanned by $\theta$ and $y$ for I, T and L fibers in Fig.~\ref{fig:fiber_trajectories}. The experiments (right panel) are obtained for a range of parameters (transversal confinement  $0.7\leq \beta\leq0.8$, lateral confinement $0.12\leq\xi\leq0.8$, and long arm aspect ratio $10\leq\ell_a/h\leq20$). In the left panel, we present the fiber trajectories obtained numerically for a single set of parameters ($\xi=0.5$, $\beta=0.76$ and $\ell_a/h = 10$) for qualitative comparison. The qualitative trends are well reproduced; differences may arise from the different values of the lateral confinement $\xi$ as will be discussed later}.\par

As already presented previously \cite{nagel2018}, I fibers show mostly permanent lateral oscillations, clockwise around the fix point at $\theta=0^\circ$ (red trajectory, Fig.~\ref{fig:reorientation}(a)) or anti-clockwise around the fix point at $\theta=\pm 90^\circ$ (green trajectory, Fig.~\ref{fig:reorientation}(b)) except if they are located very close to the wall (blue trajectory). If we reduce the level of symmetry by analyzing a T shaped fiber, the picture changes significantly. The stable fix points at the center of the channel for $\theta=0^\circ,\,\pm 90^,\,\pm 180^\circ$ are reduced to one at $\theta=0^\circ$. Apart from trajectories ending at this configuration (red trajectory, Fig.~\ref{fig:reorientation}(c)), there are also attracting points at the lateral channel walls (blue and green trajectories, {the latter corresponds to Fig.~\ref{fig:reorientation}(d)}). While approaching these final positions, the T fiber may exhibit non-monotonous lateral movement as it is rotating. Moreover, the interaction with the lateral walls can lead to a change of rotation direction, as two of the four unstable fix points observed for I fibers are still present (highlighted by red dots in Fig.~\ref{fig:fiber_trajectories}). {The green trajectory (Fig.~\ref{fig:reorientation}(d)) corresponds to an example of such a non-monotonous evolution of the orientation angle}. For the fully asymmetric L fibers, the centrally located fix points finally disappear completely and the only permanent configurations correspond to the fiber being captured by one of the walls, with the sharp edge pointing toward the channel center. Even though there exist trajectories pointing away from this configuration, they finally lead back to it and the fiber cannot escape. The stable and unstable orientations observed without lateral confinement do not appear as fix points, as they lead to lateral drift, but as ridges with the trajectories either leading away from or attracting to. In other words, after rotating toward its stable orientation, the L fiber drifts toward the wall where it will remain stuck. In some cases (light red area), the fiber is captured by the opposite wall during the reorientation. We highlight the trajectories of the three cases presented in Fig.~\ref{fig:reorientation} {(e-g)}. The fiber, initially close to the unstable orientation, leaves the orientation by rotating either clockwise (green curve, Fig. \ref{fig:reorientation} {(e)}) or anti-clockwise (red curve, Fig. \ref{fig:reorientation} {(f)}) to reach equilibrium.  When approaching the wall with the sharp edge toward it (blue curve, Fig. \ref{fig:reorientation} {(g)}{(f)}), the fiber is first repelled, rotates until reaching its equilibrium angle, and finally drifts across the channel width to reach the top wall where it will remain. Note that for a mirrored fiber of opposite chirality, there exists a mirror-symmetric diagram where most fibers collect at the opposite wall.\par

For T and L~fibers we can quantify the percentage of fibers being trapped at one of the channel walls or localized at the channel center. For T~fibers, the corresponding set of initial configurations which end up in the channel center (light green) or at the walls (light red or blue) are highlighted in Fig.~\ref{fig:fiber_trajectories}. Consequently, the areas of these regimes represent the statistical probability of finding a fiber with random initial orientation and lateral position at channel center or wall. In the presented case, approximately $40\,\%$ of the fibers end in the channel middle and the rest are captured by the lateral walls. Looking at the trajectories ending at the walls, we can see that they correspond to the case where the fiber is captured by the wall during reorientation toward the central position {(Fig.~\ref{fig:reorientation}(d))}. This means that we can tune this ratio by varying the lateral confinement $\xi$. In fact, for $\xi=0.3$, we find approximately $73\,\%$ of the fibers in the channel center. Decreasing the lateral confinement thus increases the probability of finding a T fiber in the middle of the channel. It has to be noted, however, that close to the end points at the walls, there exist trajectories which lead away again from the wall. Assuming some experimental noise, this can easily lead to a higher amount of fibers finally ending in the channel center than obtained from a noise-free theoretical study. Turning to L~fibers, most of them end at the same lateral wall (approximately $86\,\%$ for $\xi=0.5$, light blue area), which is determined by the sign of drift angle at stable orientation of a fiber without lateral confinement. Analogue to the T fiber, the partition of fibers being captured by the opposite wall can be reduced (increased) by decreasing (increasing) the lateral confinement (approx. $93\,\%$ fibers at one wall for $\xi=0.3$). See appendix \ref{app:counts} for details.

These findings show that it is possible to control the lateral position of fibers transported in confining micro-channels by tuning their geometry, in particular their symmetry properties, and/or the confinement, i.e. the channel size. I~fibers (two axes of symmetry) are mostly laterally oscillating, or tend to end in the channel center due to small imperfections (damped oscillation) \cite{nagel2018}. T~fibers (left/right symmetry, fore/aft asymmetry) are either pushed toward the channel center or toward the lateral walls, depending on the level of lateral confinement. L fibers (fully asymmetric) are always captured by the lateral walls. Depending on their chirality being left- or right-handed, most of the fibers end hereby at one or the other wall. 
{The dynamics is discussed in the appendix \ref{app:dynamics}. The reorientation time, i.e., the time needed to reach the equilibrium orientation in the absence of lateral walls, strongly depends on the transversal confinement $\beta$ but is typically $\tau\sim10-100\ell_a/U$ (see  Fig. \ref{fig:theta_evol} in appendix \ref{app:dynamics}), indicating that a fiber reaches its equilibrium position after traveling a few tens of its length. For fibers of length $\ell_a\sim 100\,\upmu$m, equilibrium is thus reached after a few centimeters. In addition, for laterally confined channels, starting from a broad initial distribution, all fibers have reached the end point of their trajectories (i.e., at the wall or at the center of the channel) after $100-1000\ell_a$ (see Fig. \ref{fig:stats} in appendix \ref{app:counts}). Typically, a channel of length $4\,$cm is sufficient for fibers to reach their final configuration (Fig. \ref{fig:reorientation}). This dynamics can be tuned by adjusting both transversal ($\beta$) and lateral ($\xi$) confinements.} The presented results thus open new routes toward the design of sorting devices or filters as a function of for example micro-organism shape. In addition they can be used to design targeted delivery or particle capture applications in microfluidic devices.


\begin{acknowledgments}
{J.C. and A.L. acknowledge the European Research Council is for funding the work through a consolidator grant (ERC PaDyFlow 682367). M.B. acknowledges the Deutsche Forschungsgemeinschaft for financial support through a research fellowship (403680998).}
\end{acknowledgments}

\appendix
\section{2D Brinkman/gap-flow model}
\label{app:Brinkman}
A detailed derivation of the reduced model can be found in \cite{nagel2018}. The flow is calculated solving the so-called Brinkman equations,
\begin{subequations}
\begin{align}
\bs\nabla\cdot\bs u &= 0,\\
\left(\nabla^2\bs u - \frac{12}{H^2}\bs u\right) - \bs \nabla p &= 0,
\end{align}
\end{subequations}
with velocity field $\bs u = (u,v)$, pressure $p$ and channel height $H$. These equations are obtained by depth-averaging the Stokes equations under the assumption of a Poiseuille-flow profile along $z$. The so-called composite particle, which consists in the particle itself and the liquid films above and below it, is treated as a rigid object in the channel, with its velocity entering as a boundary condition for the depth-averaged flow on the particle surface. The composite particle velocities in $x$- and $y$-direction as well as the rotation speed are then determined by calculating the total forces and torque resulting from four reference configurations, a resting particle in a moving surrounding and a particle moving in a quiescent fluid, either in $x$- or $y$-direction, or rotating. Due to the linearity of the Brinkman equations, it is then possible to determine the velocities corresponding to a force- and torque-free composite particle.\par

For calculation of the total force and torque, we need to specify further information on the flow in the gaps between the upper and lower particle sides and the upper and lower channel walls. Assuming a flow of Couette-Poiseuille type in the gap, with no-slip conditions on the channel walls and the particle surface, enables us to specify the force $\bs F_{\rm gap}$ and the torque $M_{\rm gap}$ on the top and bottom surface of the composite particle:
\begin{subequations}
\begin{align}
\bs F_{\rm gap} = 2\,\mu\,\bs u_p\,A_p\frac{6}{H(1-\beta^3)},\\
M_{\rm gap} = 2\,\mu\,\dot\theta_p\,T_p\frac{6}{H(1-\beta^3)},
\end{align}
\end{subequations}
with area of the top surface $A_p$, moment $T_p=\int dA\,(x^2+y^2)$, and composite particle velocity $\bs u_p$ and rotation speed $\dot\theta_p$. The assumption for the gap-flow also directly leads to a link between the composite particle velocities $(u_p,v_p,\dot\theta_p)$ and the real particle velocities,
\begin{align}
(u,v,\dot \theta) = \frac{3}{2}\frac{1+\beta}{1+\beta+\beta^2}\,(u_p,v_p,\dot\theta_p),
\end{align}\par
wherre $(u,v,\dot\theta)$ are used to update the particle position at the next time step.\par
In this work, the two-dimensional flow is solved with \textsc{comsol multiphysics} or the \textsc{ulambator} code. As inertia effects can be neglected, the particle trajectories in the $y$-$\theta$-space can simply be constructed by calculating the velocity vectors, i.e., the gradient field.

\section{Dynamics of orientation in an infinite medium ($\xi=0$)}
\label{app:dynamics}

\begin{figure}[h!]
  \begin{center}
    \includegraphics[width=0.7\columnwidth]{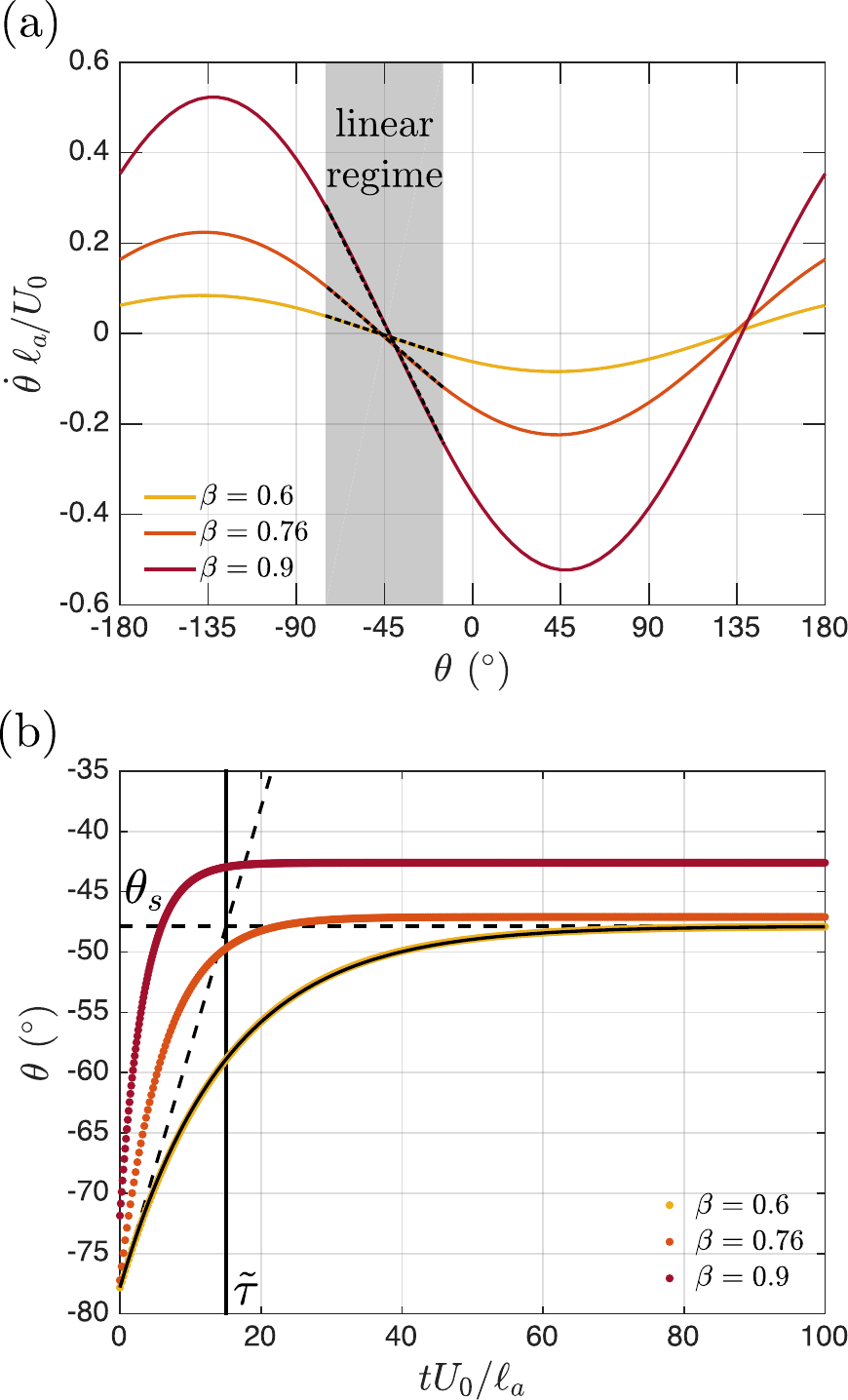}
  \end{center}
  \caption{Dynamics of orientation of an L fiber with $\ell_a/h = 10$ and $\ell_b/h = 5$  for different confinements $\beta = 0.6$ (yellow), $\beta = 0.76$ (orange) and $\beta = 0.9$ (red), as obtained from numerical calculations. (a) Evolution of the dimensionless rotation velocity $\dot{\theta} \ell_a/U_0$ as a function of the angle $\theta$ far away from the lateral walls. In the vicinity of the stable orientation angle $\theta_s$, $\dot{\theta}$ depends approximately linearly on $\theta$. (b) Evolution of the orientation angle as a function of the dimensionless time. The black curve corresponds to an exponential fit of the data using Eq.~(\ref{eq:fit_exp}). From the fit one can determine the dimensionless characteristic time of the orientation process $\tilde{\tau}$.}
\label{fig:dynamic}
\end{figure}

Figure~\ref{fig:dynamic}~(a) shows the evolution of the dimensionless rotation velocity $\dot{\theta}\,\ell_a/U_0$ as a function of the orientation angle $\theta$ for different transversal confinements.  The amplitude of $\dot{\theta}$ increases for increasing confinement, i.e., the reorientation of the fiber towards its equilibrium orientation is faster when transversal confinement increases. For orientation angles close to the stable orientation angle $\theta_s$, the evolution of $\dot{\theta}$ as a function of $\theta$ can be approximated by a linear function. The shaded region of Figure~\ref{fig:dynamic}~(a) indicates this linear regime ($| \theta-\theta_s| <30^{\circ} $), where the evolution of the orientation angle as a function of time is consequently expected to be exponential.\par 

Figure~\ref{fig:dynamic}~(b) shows the evolution of $\theta$ as a function of the dimensionless time $tU_0/\ell_a$ for three different confinements in the vicinity of the stable orientation angle $\theta_s$ as obtained from calculations with the \textsc{ulambator} code. After some time, the orientation angle reaches a plateau, which corresponds to the equilibrium orientation $\theta_s$. For increasing transversal confinement $\theta_s$ increases and the equilibrium orientation is reached faster, which can be explained by the higher rotation velocity as shown in Fig.~\ref{fig:dynamic}~(a).\par

The time evolution of the orientation angle can be fitted by an exponential saturation function:

\begin{equation}
 \theta(t U_0/h) = (\theta_{s}-\theta_0)\left(1-\exp\left(-\frac{t\,U_0}{\ell_a\tilde{\tau}}\right)\right)+\theta_0,
 \label{eq:fit_exp}
 \end{equation}
 with $\tilde{\tau}$, and $\theta_{s}$ being the two fitting parameters, and initial orientation $\theta_0$.

\begin{figure}[h!]
  \begin{center}
    \includegraphics[width=0.7\columnwidth]{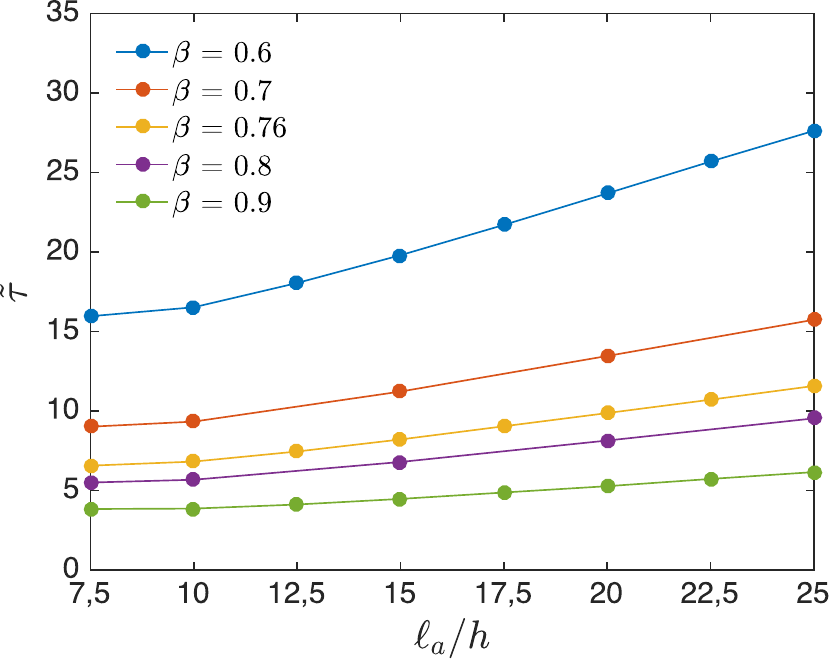}
  \end{center}
  \caption{Dimensionless characteristic time  as a function of $\ell_a/h$ for transversal confinements varying from 0.6 to 0.9. In all cases $\ell_b = 5h$.}
\label{fig:eq_t_cara}
\end{figure}

Figure~\ref{fig:dynamic}~(b) shows the fit (black curve) of the orientation angle evolution for the lowest confinement. Another example of such a fit is given in  Fig. 2~(d) of the article.
The characteristic dimensionless time $\tilde{\tau}$ characterizes the dynamics of the particle orientation and depends on the fiber lenghts $\ell_a/h$, $\ell_b/h$ and on the transversal confinement. These dependences are shown in Fig.~\ref{fig:eq_t_cara}, where $\ell_b = 5h$ in order to stay consistent with the experiment. For both increasing confinement and increasing $\ell_a$, $\tilde{\tau}$ increases as well. \\

\section{Experimental measure of the equilibrium angle ($\xi=0$)}
\label{app:exp_measure}

Experimentally, the steady orientation angle of a L fiber evolving in an infinitely wide channel ($\xi = 0$) is determined using intermediate numerical results. The measuring protocol is described in the following.
\begin{figure}[h!]
  \begin{center}
    \includegraphics[width=0.7\columnwidth]{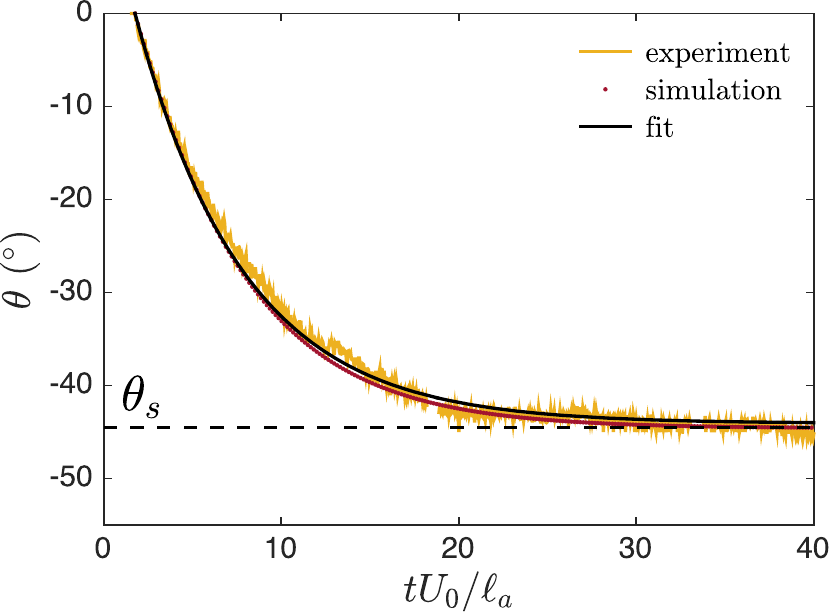}
  \end{center}
  \caption{Evolution of the orientation angle as a function of the dimensionless time for an L fiber with $\ell_a/h = 7.5$, $\ell_b/h = 5$, and $\beta=0.76$. A very good agreement is obtained between experiments (yellow), simulations (red), and the exponential fit (black).}
\label{fig:theta_evol}
\end{figure}
For short fibers ($\ell_a/h<12.5$ and $\ell_b/h=5$) the influence of the lateral walls is negligible and the transient evolution of the orientation angle exhibits as expected an exponential evolution toward a plateau. Figure~\ref{fig:theta_evol} shows the superposition of the this evolution as obtained experimentally (yellow curve) and in the simulation (red dots), as well as the exponential fit (black curve). The three curves show very good agreement validating both the numerical results and the pertinence of the exponential fit.\par

For larger fibers, the orientation angle never reaches a plateau, which makes a direct measurement of the equilibrium angle impossible. Moreover, the fiber reorients with a longer characteristic time as compared to shorter fibers and drifts toward a lateral wall which impact the fiber trajectory and the orientation angle evolution. However, knowing the dimensionless characteristic time $\tilde{\tau}$ one can extrapolate the experimental observation of the orientation angle evolution. To distinguish the dynamics caused solely by to the particle shape and the external flow from those caused by an interaction with the wall, we fit a portion of the transient evolution of $\theta$, corresponding to the situation where the fibers is far from the walls, by the exponential function (\ref{eq:fit_exp}). We keep $\theta_{s}$ as a fitting parameter but set the value of $\tilde{\tau}$ to the one obtained numerically. This one-parameter fitting method allows us to extract the steady orientation angle from the experiments. Note that this procedure is based on the good agreement between numerical results and experiments (see Fig.~\ref{fig:theta_evol}) not only with respect to the equilibrium angle, but also regarding the transient evolution, i.e., the characteristic time of reorientation $\tilde \tau$.

\section{Effect of lateral confinement on the fiber distribution in the channel}
\label{app:counts}
\begin{figure}
  \begin{center}
    \includegraphics[width=1\columnwidth]{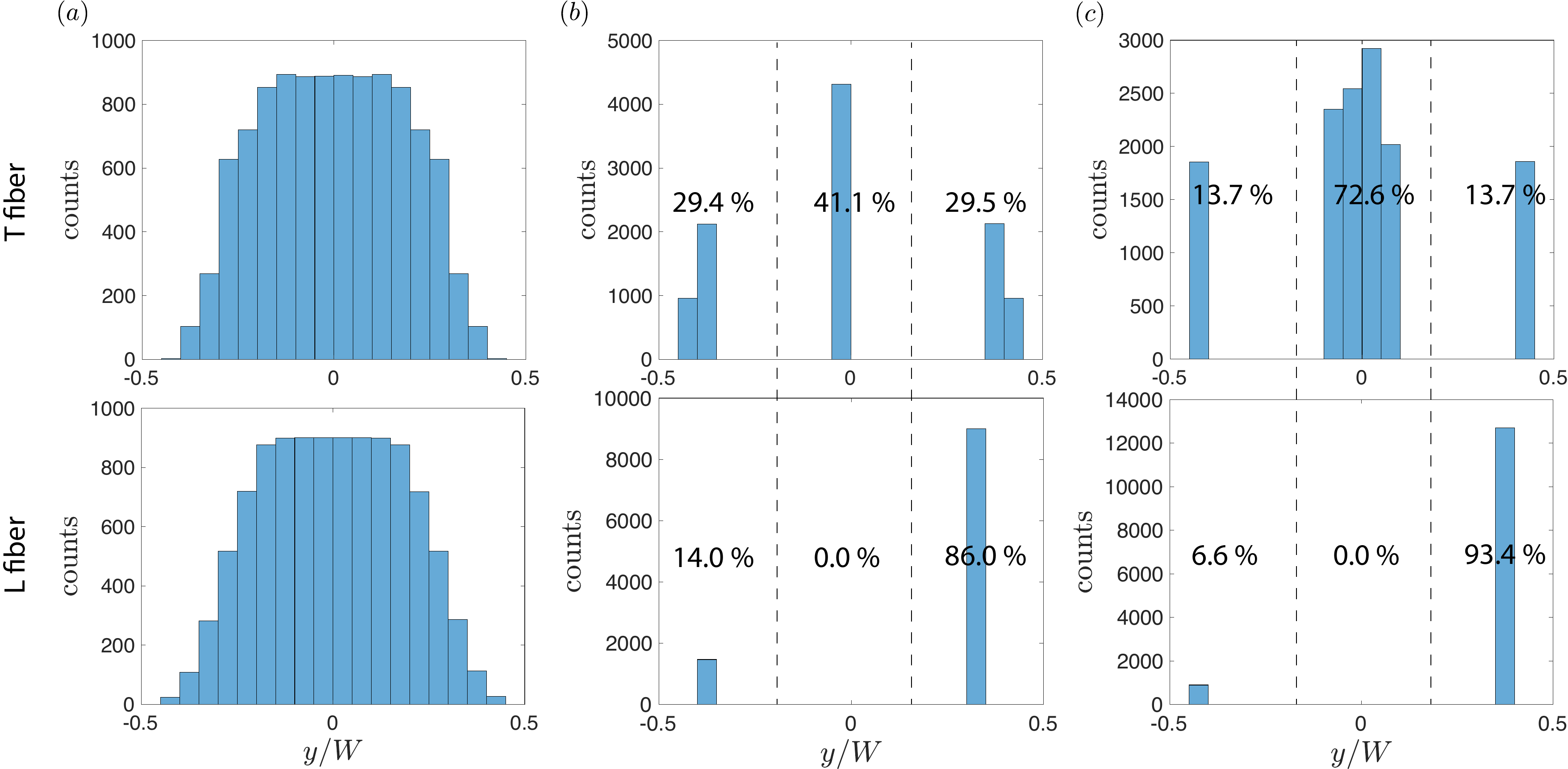}
  \end{center}
  \caption{Distribution of lateral position in the channel for T (top) and L (bottom) fibers: Initial distribution for $\xi = 0.5$ (a), and distribution after $L=10^3\,\ell_a$ for (b) $\xi=0.5$ and (c) $\xi=0.3$.}
\label{fig:stats}
\end{figure}

We calculate the trajectories of T and L fibers for at least $10^4$ initial configurations equally distributed in the configuration space. Figure~\ref{fig:stats}(a) shows the distribution in lateral position for a lateral confinement of $\xi = 0.5$. We then analyze the distribution after transportation in a channel of length $L=10^3\,\ell_a$. As shown by Fig.~\ref{fig:stats}(b), T fiber are either collected in the channel center ($\sim 41\%$) or at both lateral walls ($\sim 59\%$), while most of the L fibers accumulate at one of the lateral walls ($86\%$). This effect can be tuned by changing the lateral confinement, as visualized by Fig.~\ref{fig:stats}(c) for $\xi = 0.3$. In this case, more T fibers are at the channel center ($\sim 73\%$) and the percentage of L fibers being located at the same lateral wall is increased.


\begin{thebibliography}{10}

\bibitem{Gossett2010}
D.~R. Gossett, W.~M. Weaver, A.~J. Mach, S.~C. Hur, H.~T.~K. Tse, W.~Lee,
  H.~Amini, and D.~Di~Carlo, ``Label-free cell separation and sorting in
  microfluidic systems,'' {\em Analytical and bioanalytical chemistry},
  vol.~397, no.~8, pp.~3249--3267, 2010.

\bibitem{DiCarlo2009}
D.~Di~Carlo, ``Inertial microfluidics,'' {\em Lab on a Chip}, vol.~9, no.~21,
  pp.~3038--3046, 2009.

\bibitem{Masaeli2012}
M.~Masaeli, E.~Sollier, H.~Amini, W.~Mao, K.~Camacho, N.~Doshi, S.~Mitragotri,
  A.~Alexeev, and D.~Di~Carlo, ``Continuous inertial focusing and separation of
  particles by shape,'' {\em Physical Review X}, vol.~2, no.~3, p.~031017,
  2012.

\bibitem{DAvino2017}
G.~D'Avino, F.~Greco, and P.~L. Maffettone, ``Particle migration due to
  viscoelasticity of the suspending liquid and its relevance in microfluidic
  devices,'' {\em Annual Review of Fluid Mechanics}, vol.~49, pp.~341--360,
  2017.

\bibitem{Secomb2017}
T.~W. Secomb, ``Blood flow in the microcirculation,'' {\em Annual Review of
  Fluid Mechanics}, vol.~49, pp.~443--461, 2017.

\bibitem{duRoure2019}
O.~du~Roure, A.~Lindner, E.~N. Nazockdast, and M.~J. Shelley, ``Dynamics of
  flexible fibers in viscous flows and fluids,'' {\em Annual Review of Fluid
  Mechanics}, vol.~51, pp.~539--572, 2019.

\bibitem{TenHagen2014}
B.~ten Hagen, F.~K{\"u}mmel, R.~Wittkowski, D.~Takagi, H.~L{\"o}wen, and
  C.~Bechinger, ``Gravitaxis of asymmetric self-propelled colloidal
  particles,'' {\em Nature Communications}, vol.~5, pp.~4829 EP --, 09 2014.

\bibitem{liebchen2018}
B.~Liebchen, P.~Monderkamp, B.~ten Hagen, and H.~L{\"o}wen, ``Viscotaxis:
  microswimmer navigation in viscosity gradients,'' {\em Physical review
  letters}, vol.~120, no.~20, p.~208002, 2018.

\bibitem{Marcos2009}
H.~C. Fu, T.~R. Powers, R.~Stocker, {\em et~al.}, ``Separation of microscale
  chiral objects by shear flow,'' {\em Physical review letters}, vol.~102,
  no.~15, p.~158103, 2009.

\bibitem{nagel2018}
M.~Nagel, P.-T. Brun, H.~Berthet, A.~Lindner, F.~Gallaire, and C.~Duprat,
  ``Oscillations of confined fibres transported in microchannels,'' {\em
  Journal of Fluid Mechanics}, vol.~835, pp.~444--470, 2018.

\bibitem{Berthet2013}
H.~Berthet, M.~Fermigier, and A.~Lindner, ``{Single fiber transport in a
  confined channel: Microfluidic experiments and numerical study},'' {\em
  Physics of Fluids}, vol.~25, no.~10, 2013.

\bibitem{uspal2013}
W.~E. Uspal, H.~B. Eral, and P.~S. Doyle, ``Engineering particle trajectories
  in microfluidic flows using particle shape,'' {\em Nature Communications},
  vol.~4, 2013.

\bibitem{Bet2018b}
B.~Bet, R.~Georgiev, W.~Uspal, H.~B. Eral, R.~V. Roij, and S.~Samin,
  ``{Calculating the motion of highly confined , arbitrary ? shaped particles
  in Hele ? Shaw channels},'' {\em Microfluidics and Nanofluidics}, vol.~22,
  no.~8, pp.~1--12, 2018.

\bibitem{shen2014}
B.~Shen, M.~Leman, M.~Reyssat, and P.~Tabeling, ``Dynamics of a small number of
  droplets in microfluidic hele--shaw cells,'' {\em Experiments in Fluids},
  vol.~55, no.~5, p.~1728, 2014.

\bibitem{Bet2018}
B.~Bet, S.~Samin, R.~Georgiev, H.~B. Eral, and R.~v. Roij, ``Steering particles
  by breaking symmetries,'' {\em Journal of Physics: Condensed Matter},
  vol.~30, no.~22, p.~224002, 2018.

\bibitem{Cappello2019}
J.~Cappello, M.~Bechert, C.~Duprat, O.~Du~Roure, F.~Gallaire, and A.~Lindner,
  ``Transport of flexible fibers in confined microchannels,'' {\em Physical
  Review Fluids}, vol.~4, no.~3, p.~034202, 2019.

\bibitem{Cox1970}
R.~G. Cox, ``The motion of long slender bodies in a viscous fluid {Part} 1.
  {General} theory,'' {\em Journal of Fluid Mechanics}, vol.~44, no.~4,
  pp.~791--810, 1970.

\bibitem{Candelier2016}
F.~Candelier and B.~Mehlig, ``Settling of an asymmetric dumbbell in a quiescent
  fluid,'' {\em Journal of Fluid Mechanics}, vol.~802, pp.~174--185, 2016.

\bibitem{Tozzi2011}
E.~J. Tozzi, C.~T. Scott, D.~Vahey, and D.~J. Klingenberg, ``{Settling dynamics
  of asymmetric rigid fibers},'' {\em Physics of Fluids}, vol.~23, no.~3,
  pp.~1--10, 2011.

\bibitem{Palusa2018}
M.~Palusa, J.~de~Graaf, A.~Brown, and A.~Morozov, ``Sedimentation of a rigid
  helix in viscous media,'' {\em Physical Review Fluids}, vol.~3, no.~12,
  p.~124301, 2018.

\bibitem{Dendukuri2007}
D.~Dendukuri, S.~S. Gu, D.~C. Pregibon, T.~A. Hatton, and P.~S. Doyle,
  ``{Stop-flow lithography in a microfluidic device},'' {\em Lab on a Chip},
  vol.~7, no.~7, p.~818, 2007.

\bibitem{Dendukuri2008}
D.~Dendukuri, P.~Panda, R.~Haghgooie, J.~M. Kim, T.~A. Hatton, and P.~S. Doyle,
  ``{Modeling of oxygen-inhibited free radical photopolymerization in a PDMS
  microfluidic device},'' {\em Macromolecules}, vol.~41, no.~22,
  pp.~8547--8556, 2008.

\bibitem{Nagel2015}
M.~Nagel and F.~Gallaire, ``{Boundary elements method for microfluidic
  two-phase flows in shallow channels},'' {\em Computers and Fluids}, vol.~107,
  pp.~272--284, 2015.

\end{thebibliography}

\end{document}